\documentclass[12pt,a4paper,final]{iopart}

\usepackage{iopams}  
\usepackage{graphicx}
\usepackage[breaklinks=true,colorlinks=true,linkcolor=blue,urlcolor=blue,citecolor=blue]{hyperref}

\usepackage{braket}
\expandafter\let\csname equation*\endcsname\relax
\expandafter\let\csname endequation*\endcsname\relax
\usepackage{amsmath,amsfonts,amssymb,mathrsfs }
\usepackage[toc,page]{appendix}
\usepackage{varwidth}
\allowdisplaybreaks
\usepackage[usenames]{color}
\usepackage{multirow}
\usepackage{array}
\usepackage{physics}
\usepackage[utf8]{inputenc}
\usepackage{float,diagbox}
\usepackage{varwidth}
\usepackage{subcaption}
\usepackage{makecell}

\begin{document}

\newcommand{\RW}[1]{{\color{green}[RW: #1]}}
\def\tA{\tilde{A}}
\def\bG{\mathbb{G}}
\newcommand{\hA}{\mathcal{H}_A}
\newcommand{\hB}{\mathcal{H}_B}
\newcommand{\hlA}{\mathcal{H}_{\tilde A}}
\newcommand{\tlA}{\tilde{A}}
\newcommand{\tlB}{\tilde{B}}
\newcommand{\tlC}{\tilde{C}}
\newcommand{\tla}{\tilde{a}}
\newcommand{\bIB}{\mathbb{I}_B}
\newcommand{\bIA}{\mathbb{I}_A}
\newcommand{\negaAB}{negativity susceptibility }
\newcommand{\negaABshort}{S}
\newcommand{\ot}{\otimes}
\newcommand*{\dt}[1]{%
  \accentset{\mbox{\large\bfseries .}}{#1}}
\newcommand*{\ddt}[1]{%
  \accentset{\mbox{\large\bfseries .\hspace{-0.25ex}.}}{#1}}

\def\pra{Phys. Rev. A}
\def\prl{Phys. Rev. Lett.}

\title[Separable Ball Around Any Full-Rank Multipartite Product State]{Separable Ball around any Full-Rank Multipartite Product State}

\author{Robin Yunfei Wen$^{1}$ and Achim Kempf$^{2,3,4}$} 

\address{$^1$California Institute of Technology, 1200 E. California Boulevard, Pasadena, CA 91125, USA}

\address{$^2$Department of Applied Mathematics, University of Waterloo, Waterloo, Ontario, N2L 3G1, Canada}

\address{$^3$Department of Physics, University of Waterloo, Waterloo, Ontario, N2L 3G1, Canada}

\address{$^4$Institute for Quantum Computing, University of Waterloo, Waterloo, Ontario, N2L 3G1, Canada}
\ead{ywen@caltech.edu}

\begin{abstract}
We show that around any $m$-partite product state $\rho_{\rm prod}=\rho_1\ot...\ot\rho_m$ of full rank (that is ${\rm det}(\rho_{\rm prod})\neq 0)$, there exists a finite-sized closed ball of separable states centered around $\rho_{\rm prod}$ whose radius is $\beta:=2^{1-m/2}\lambda_{\rm min}(\rho_{\rm prod})$. Here, $\lambda_{\rm min}(\rho_{\rm prod})$ is the smallest eigenvalue of $\rho_{\rm prod}$.  We are assuming that the total Hilbert space is finite dimensional and we use the notion of distance induced by the Frobenius norm. Applying a scaling relation, we also give a new and simple sufficient criterion for multipartite separability based on trace: $\Tr[\rho\rho_{\rm prod}]^2/\Tr[\rho^2]\geq \Tr\small[\rho_{\rm prod}^2\small]-\beta^2$. Using the separable balls around the full-rank product states, we discuss the existence and possible sizes of separable balls around any multipartite separable states, which are important features for the set of all separable states. We discuss the implication of these separable balls on entanglement dynamics. 
\end{abstract}

\section{Introduction}

Entanglement plays a crucial role in quantum technologies, including quantum computing \cite{10NielsenChuangText,2010LaddQCReview,17BoyerDQC1,18KlcoDigitizedSFQC}, quantum communication \cite{10YuanPhotonQCCommR,2021ChenQComm,18StephanieQI} and quantum sensing \cite{17DegenQS,17RosskpfQS,16HuntemannIonClock}. An important problem, therefore, is to obtain a quantitative understanding of the distribution of the occurrence or absence of entanglement among quantum states. 

Here, we investigate this problem for systems composed of $m$ subsystems whose Hilbert spaces, $\mathcal{H}_{i}$, $i=1,...,m$ are finite dimensional. We start by recalling that a state $\rho$ acting on $\mathcal{H}_{1}\otimes...\otimes\mathcal{H}_{m}$ is called unentangled, i.e., separable, if and only if there exist density matrices $\{\rho_{i,k}\}$ which act on the $\mathcal{H}_{i}$ respectively, such that $\rho=\sum_{k}{p_k}\rho_{1,k}\ot...\ot\rho_{m,k}, \text{ where }\sum_{k}{p_k}=1 \text{ and } p_k\geq 0 \text{ for all } k.$ 

Despite the apparent simplicity of the condition of separability, determining whether a given state, even of only two quantum systems, is entangled or separable is in general difficult and believed to be NP-hard \cite{03Gurvitis_Complexity,08GharibianNP}. Considerable efforts
have been expended on developing various methods for detecting entanglement and characterizing separable states, including the Peres-Horodecki criterion \cite{96PeresSeparability,Horodecki1996}, the entanglement witness approach \cite{00Terhal_Witness}, the development of geometric measures of entanglement \cite{01KusGeometry}, as well as the concept of entanglement monotone \cite{00EMproposal,02EMNegativity}. The Peres-Horodecki criterion is both necessary and sufficient for separability of bipartite states with $2\times 2$ or $2\times 3$ dimensions \cite{96HorodeckiSeparability}. However, no simple criterion that is both necessary and sufficient for separability in bipartite quantum systems with arbitrary finite dimension has been found to date \cite{06ZyczkowskiReview}. The separability problem in the multipartite case is even more challenging \cite{19CunhaTripartiteEntanglement,16BengtssonMulti-partite}.

A particularly interesting aspect of the separability problem is the characterization of separable balls in the space of  quantum states. As is easily seen, the set of all separable states is convex, and a key aspect of the geometry of a convex set
is the size of the largest ball that fits inside the convex set. For the maximally mixed bipartite state $\frac{1}{d} \mathbb{I}_1\otimes \mathbb{I}_2$ where $d$ is the total dimension of the system, Ref.~\cite{98Zyczkowski_Separable_Ball,99Vidal,99Braunstein_Separable_Mutipartite} showed the existence of a separable ball around the maximally mixed bipartite state, while Ref.~\cite{02GurvitsLSBaMMBS} found the exact size of the largest separable ball in the Frobenius norm as well as other spectral $l_p$ norms for $1\leq p\leq \infty$. Ref.~\cite{15ShenLargerBall} further improved upon the results of \cite{02GurvitsLSBaMMBS} through finding a slightly larger separable ball around $\frac{1}{d} \mathbb{I}_1\otimes \mathbb{I}_2$ by using the nested matrix norm $||.||_{\infty,\infty}$, which is tighter than the Frobenius norm. 

Moving beyond the bipartite case, Ref.~\cite{99Braunstein_Separable_Mutipartite,01Rungta_Qudit,03Gurvits_Multipartite_Mixed} showed the existence of a separable ball around the maximally mixed multipartite states $\frac{1}{d} \mathbb{I}_1\otimes...\otimes \mathbb{I}_m$ while providing lower and upper bounds to the radius of the largest separable ball. To our knowledge, no exact radius has been found yet for the largest separable ball around the the maximally mixed multipartite state in the general case with arbitrary dimensions and arbitrary numbers of systems. Despite these results on separable balls around the maximally mixed state, much less is known about the existence and possible sizes of separable balls around other bipartite or multipartite states of interest, such as the product states of the form $\rho_1\otimes...\otimes\rho_m$.

In this context, Ref.~\cite{16LamiEntanglementSaving} showed the existence of a separable ball around any full-rank bipartite product state $\rho_1\otimes\rho_2$ with ${\rm det}(\rho_1\otimes\rho_2)>0$, while for any $\rho_1\otimes\rho_2$ with ${\rm det}(\rho_1\otimes\rho_2)=0$, it lies on $\partial S$, the boundary of the set of all separable states $S$. Ref.~\cite{20Hanson_Markovian} further strengthens the result by finding a lower bound on the radius of the separable ball to be $\lambda_{\rm min}(\rho_1)\lambda_{\rm min}(\rho_2)$ in the bipartite case. In the present work, we improve upon the results of \cite{16LamiEntanglementSaving,20Hanson_Markovian} by considering the general case of $m$-partite systems in Sec.~\ref{sec:prod-ball}. We prove that there always exists such a separable ball around any full-rank $\rho_{\rm prod}=\rho_1\ot...\ot\rho_m$ (${\rm det}(\rho_{\rm prod})>0$). We also find a lower bound for the radius of the ball to be $2^{1-m/2}\lambda_{\rm min}(\rho_{\rm prod})$, which is proportional to the smallest eigenvalue of $\rho_{\rm prod}$ and exponentially decreasing with respect to the number of systems $m$. Our work has been partially inspired by \cite{21Qian}, which developed a numerical algorithm for applying the bipartite separability criterion in \cite{98Zyczkowski_Separable_Ball,02GurvitsLSBaMMBS} to states that are not in the vicinity of the maximally mixed states $\frac{1}{d} \mathbb{I}_1\otimes\mathbb{I}_2$ by using transformations of the form $(X\ot Y)B(X\ot Y)^{\dag}$ such that $X$ and $Y$ are invertible.

Applying a scaling relation to the result in Sec.~\ref{sec:prod-ball}, we also give a new, simple sufficient criterion for multipartite separability based on trace in Sec.~\ref{sec:criterion}: $\Tr[\rho\rho_{\rm prod}]^2/\Tr[\rho^2]\geq \Tr[\rho_{\rm prod}^2]-\beta^2$, where $\beta:=2^{1-m/2}\lambda_{\rm min}(\rho_{\rm prod})$. Using the separable balls around the full-rank product states, we discuss the existence and
possible sizes of separable balls around any multipartite separable states in Sec.~\ref{sec:general-ball}, which are
important features for the interior and the boundary of the set of all separable states. Last, we discuss the implication of these separable balls on the entanglement dynamics in Sec.~\ref{sec:dynamics}, which strengthens and generalizes the results of Section 3 in \cite{22Wen}.

\section{Main Result: A separable ball exists around any positive-definite, Hermitian \texorpdfstring{\boldmath{$A_1\otimes...\otimes A_m$}}{A1Am}} \label{sec:proof}\label{sec:prod-ball}

Let us recall that the usual definition of the separability of density matrices trivially generalizes to Hermitian matrices. 
In the following, it will be convenient to work initially with Hermitian matrices and then to specialize to density matrices. We begin by stating and proving the main result of the present work: \medskip\newline
\noindent
{\bf Theorem 1}: Consider arbitrary positive definite Hermitian operators $A_i$, $1\leq i\leq m$, acting on finite-dimensional Hilbert spaces $\mathcal{H}_{i}$. We define $A_{\rm prod}:=A_1\ot...\ot A_m$. Then if a Hermitian operator $B$ acting on $\mathcal{H}_{1}\otimes...\otimes \mathcal{H}_{m}$ satisfies $||B-A_{\rm prod}||_F\leq 2^{1-m/2}\lambda_{\rm min}(A_{\rm prod})$, then $B$ is separable. Here, $||X||_F=\sqrt{\Tr[X^{\dagger}X]}$ is the Frobenius norm and $\lambda_{\rm min}(X)$ is the smallest eigenvalue of $X$. \\
\\
To prove Theorem 1, we make use of the following three results:\medskip\newline
\noindent
{\bf Result 1}: In a finite-dimensional $m$-partite system $\mathcal{H}_{1}\otimes...\otimes\mathcal{H}_{m}$, the Hermitian matrix $B$ is separable if $||B-\mathbb{I}||_F\leq 2^{1-m/2}$, where $\mathbb{I}$ is the identity matrix.\medskip\newline
{\bf Result 2}: If a Hermitian matrix $B$ acting on $\mathcal{H}_{1}\otimes...\otimes\mathcal{H}_{m}$ is separable, then $(X_1\ot...\ot X_m)B(X_1\ot...\ot X_m)^{\dag}$ is separable for any invertible $X_i$ acting on $\mathcal{H}_{i}$ for all $1\leq i\leq m$.\medskip\newline
{\bf Result 3}: For any diagonal matrix $D$ and an arbitrary matrix $A$, we have $||DAD||_F\leq {\rm max}(D_{ii}^2)||A||_F$.
\medskip\newline
Result 1 is stated and proven as Corollary 1 in \cite{03Gurvits_Multipartite_Mixed}, and it is crucial to proving the existence of separable balls around the maximally mixed $m$-partite states in \cite{03Gurvits_Multipartite_Mixed}. In the bipartite case with $m=2$, the bound in Result 1 becomes $2^{1-m/2}=1$, which reduces to Theorem 1 in \cite{02GurvitsLSBaMMBS}. The bipartite version of Result 2 is stated and discussed as Lemma 1 in \cite{21Qian}, and the full multipartite case is straightforward to show with the definition of separability for Hermitian matrices. We here only prove Result 3, which is critical to this work for finding the radius of the separable ball around $A_1\otimes...\otimes A_m$ in the Frobenius norm.\\
\noindent \bf Proof of Result 3: \rm We know that $(DAD)_{ij}=\sum_{l,m}D_{il}A_{lm}D_{mj}=D_{ii}D_{jj}A_{ij}$ since $D$ is diagonal. Therefore,
\begin{align}
    ||DAD||_F^2&=\sum_{i,j}|(DAD)_{ij}|^2=\sum_{i,j}\left \lvert D_{ii}D_{jj}\right\rvert^2|A_{ij}|^2\nonumber\\
    &\leq  {\rm max}\{|D_{ii}D_{jj}|^2:\forall i,j\} \sum_{i,j}|A_{ij}|^2={\rm max}\{|D_{ii}|^2\}^2||A||_F^2,
\end{align}
so clearly $||DAD||_F\leq {\rm max}(D_{ii}^2)||A||_F$. The above equality is achieved when $|D|$ has the same value across the entire diagonal.\\
\\
{\bf Proof of Theorem 1:} We can work in a basis of $\mathcal{H}_1\otimes...\otimes \mathcal{H}_m$ such that all $A_i$ are diagonal and positive-definite, since the Frobenius norm is invariant under unitary transformations. Let
\begin{align}
    C&:=B-A_{\rm prod},\\
    \Delta&:= A_{\rm prod}^{-\frac{1}{2}}C A_{\rm prod}^{-\frac{1}{2}} \label{eq:Delta}.
\end{align}
From the theorem's assumption, we know that $||C||_F\leq 2^{1-m/2}\lambda_{\rm min}(A_{\rm prod})$. Clearly, $A_{\rm prod}$ is diagonal and invertible. Applying Result 3, we then have
\begin{align}
    ||\Delta||_F&= \left\lvert\left\lvert A_{\rm prod}^{-\frac{1}{2}}C A_{\rm prod}^{-\frac{1}{2}}\right\rvert\right\rvert_F \leq  {\rm max} \left\{(A_{\rm prod}^{-\frac{1}{2}})_{ii}^2\right\}||C||_F={\rm max}\left\{(A_{\rm prod}^{-1})_{ii}\right\}||C||_F\nonumber\\
    &=\frac{1}{\lambda_{\rm min}(A_{\rm prod})}||C||_F=2^{1-m/2}
\end{align}
As defined in Eq.~\ref{eq:Delta}, $\Delta$ is indeed Hermitian since $A$ is Hermitian, so $\mathbb{I}+\Delta$ is Hermitian. Through Result 1, we know that $\mathbb{I}+\Delta$ is separable. Using Result 2 where the product state $A_{\rm prod}^{\frac{1}{2}}$ is invertible, we have
\begin{align}
    A_{\rm prod}^{\frac{1}{2}}(\mathbb{I}+\Delta)A_{\rm prod}^{\frac{1}{2}}=A_{\rm prod}+C=B
\end{align}
is also separable, which completes the proof. 
\medskip\newline\noindent
Notice that Theorem 1 can be viewed as a generalization of Result 1. When the $A_i$ in Theorem 1 are all identity matrices, then $\lambda_{\rm min}(\mathbb{I})=1$, reducing to Result 1. In the bipartite case with $m=2$, the radius becomes $\lambda_{\rm min}(A_1\ot A_2)$, which recovers Lemma 2.2 of \cite{20Hanson_Markovian}. The radius $\lambda_{\rm min}(A_1\ot A_2)$ for the separable around $A_1\otimes A_2$ is in fact the largest in the space of Hermitian matrices. Consider $A_1\ot A_2-a\ket{x}\bra{x}$ where $\ket{x}$ is the eigenvector corresponding to $\lambda_{\rm min}(A_1\otimes A_2)$. The matrix becomes negative if $a>\lambda_{\rm min}(A_1\ot A_2)$, and it will no longer be separable. Such a construction does not work for the multipartite case due to the exponential decrease in the $2^{1-m/2}$ term, so our radius is generally not the largest in the space of Hermitian matrices for $m\geq 3$.

We next extend the radius in Theorem~1 from the Frobenius norm (Schatten 2-norm) to other Schatten $p-$norms by using common inequalities in $l_p$ norms. For a Hermitian matrix $\Delta$, the Schatten $p-$norm is defined as $||\Delta||_p:=\Tr[|\Delta|^p]^{\frac{1}{p}}=(\sum_i |\lambda_i(\Delta)|^p|)^{\frac{1}{p}}$, where $\lambda_i(\Delta)$ are eigenvalues of $\Delta$. Since $||\Delta||_2\leq||\Delta||_p$ for $1\leq p\leq 2$, a lower bound on the radius of the separable ball around $A_{\rm prod}$ is also $2^{1-m/2}\lambda_{\rm min}(A_{\rm prod})$ in the metric induced by the Schatten $p-$norm with $1\leq p\leq 2$. With $||\Delta||_2\leq d^{\frac{1}{2}-\frac{1}{p}}||\Delta||_p$ for $2\leq p\leq \infty$ where $d$ is the dimension of the matrix $\Delta$, we can conclude that when $||B-A_{\rm prod}||_p\leq d^{\frac{1}{p}-\frac{1}{2}}2^{1-m/2}\lambda_{\rm min}(A_{\rm prod})$ for $2\leq p\leq \infty$, then $B$ is separable. 

For density matrices which describe physical quantum states, Theorem~1 can be applied to any multipartite product state $\rho_{\rm prod}=\rho_1\ot...\ot\rho_m$ that is fully mixed (or full-rank), in the sense that $\rho_{\rm prod}$ has no vanishing eigenvalue (i.e., if it is positive definite). Our main finding is that there exists a finite-sized separable ball around any full-rank multipartite product state $\rho_{\rm prod}$ with radius $\beta:=2^{1-m/2}\lambda_{\rm min}(\rho_{\rm prod})=2^{1-m/2}\prod_{i=1}^m\lambda_{\rm min}(\rho_i)$ in the Frobenius norm. For the bipartite, positive-definite product state, the radius of the separable around $\rho_1\otimes\rho_2$ is simply $\lambda_{\rm min}(\rho_1)\lambda_{\rm min}(\rho_2)$, which recovers the result in \cite{20Hanson_Markovian}. Even though our radius $\lambda_{\rm min}(\rho_1 \otimes \rho_2)$ for the bipartite separable ball is the largest in the space of Hermitian matrices, it is generally not the largest in the space of density matrices. This can be seen in the case $\rho_1\otimes\rho_2=\frac{1}{d}I$, where the largest radius of the separable ball for the maximally mixed bipartite state was found to be $r_{\rm max}=1/\sqrt{d(d-1)}$ in \cite{02GurvitsLSBaMMBS}, while our result here gives $r=1/d$, which is slightly smaller than $r_{\rm max}$. Ref.~\cite{02GurvitsLSBaMMBS} derives $r_{\rm max}$ by using a scaling relation and the fact that $\Tr[\rho]=1$ along with Result 1. We will use a similar scaling relation to improve Theorem~1 in Sec.~\ref{sec:criterion}, although the improved separability criterion will no longer possess the geometry of a ball. 

We notice that the radius of the separable ball $\beta=2^{1-m/2}\lambda_{\rm min}(\rho_{\rm prod})$ decreases exponentially with respective to the number of particle $m$, which is consistent with the numerical observation in \cite{98Zyczkowski_Separable_Ball} that the volume of the set of all separable states decreases exponentially with the dimension of the Hilbert space. We also make the comment that the separable balls around the full-rank bipartite product states are generally not absolutely separable, even though the separable ball around the maximally mixed state $\frac{1}{d}\mathbb{I}$ is \cite{14Arunachalam_AS}. A quantum state $\rho$ is defined to be absolutely separable when $U\rho U^{\dagger}$ is separable for all unitary matrices $U$ acting on $\mathcal{H}_1\otimes\mathcal{H}_2$. In fact, there exists full-rank product states that are not absolutely separable. One numerical example is given in Figure~2 of \cite{22Wen}, where a full-rank $\rho_A\otimes\rho_B$ becomes entangled under unitary evolution, so the state is not absolutely separable. 

In the space of density matrices, we can also write the separable ball around $\rho_{\rm prod}$ in terms of the quantum relative entropy $D(\rho||\sigma):=\Tr[\rho(\log\rho-\log\sigma)]$. Using the Pinsker's Inequality $\frac{1}{2}||\rho-\sigma||_1^2\leq D(\rho||\sigma)$ \cite{Pinsker}, we have if $D(\rho||\rho_{\rm prod})\leq 2^{1-m}\lambda_{\rm min}^2(\rho_{\rm prod})$, then $||\rho-\rho_{\rm prod}||_F\leq||\rho-\rho_{\rm prod}||_1\leq 2^{1-m/2}\lambda_{\rm min}(\rho_{\rm prod})$, so $\rho$ is separable.

\section{A Sufficient Separability Criterion: \texorpdfstring{\boldmath{$\Tr[\rho\rho_{\rm prod}]^2/\Tr[\rho^2]\geq \Tr\small[\rho_{\rm prod}^2\small]-\beta^2$}}{crit}}\label{sec:criterion}

As we discussed, the radius of the separable balls around $\rho_{\rm prod}$ given by Theorem~1 is generally not the largest. Here we aim to improve on the condition $||\rho-\rho_{\rm prod}||\leq 2^{1-m/2}\lambda_{\rm min}(\rho_{\rm prod}):=\beta$ by finding a slightly larger subset of $\rho$ around $\rho_{\rm prod}$ that is separable. To show $\rho$ is separable is equivalent to showing $\alpha\rho$ is separable for some $\alpha>0$. If we can find any $\alpha>0$ such that $\alpha\rho=\rho_{\rm prod}+\Delta$ where $||\Delta||_F\leq \beta$, then $\rho$ will be separable. With
\begin{align}
    ||\Delta||_F^2&=\Tr[\rho^2]\alpha^2-2\Tr[\rho\rho_{\rm prod}]\alpha+\Tr[\rho_{\rm prod}^2]\label{eq:Frobenius-scaling},
\end{align}
which is quadratic with respect to $\alpha$, then we can take $\alpha=\Tr[\rho\rho_{\rm prod}]/\Tr[\rho^2]>0$ to minimize $||\Delta||_F^2$. Minimizing $||\Delta||_F^2$ will ensure a larger separable subset around $\rho_{\rm prod}$. If 
\begin{align}
||\Delta||_F^2&=\Tr[\rho_{\rm prod}^2]-\frac{\Tr[\rho\rho_{\rm prod}]^2}{\Tr[\rho^2]}\leq\beta^2
\Longleftrightarrow \frac{\Tr[\rho\rho_{\rm prod}]^2}{\Tr[\rho^2]}\geq\Tr[\rho_{\rm prod}^2]-\beta^2
    \label{eq:condition-2},
\end{align}
then $\rho$ is separable. To show that Eq.~\ref{eq:condition-2} is indeed a more general sufficient condition than the separable ball criterion $||\rho-\rho_{\rm prod}||_F\leq \beta$, we see that Eq.~\ref{eq:condition-2} is equivalent to
\begin{align}
||\rho-\rho_{\rm prod}||_F^2&=\Tr[\rho^2]-2\Tr[\rho\rho_{\rm prod}]+\Tr[\rho_{\rm prod}^2]\nonumber\\
&\leq \frac{\Tr[\rho\rho_{\rm prod}]^2}{\Tr[\rho_{\rm prod}^2]-\beta^2}-2\Tr[\rho\rho_{\rm prod}]+\Tr[\rho_{\rm prod}^2]\nonumber\\
&=\beta^2+\frac{1}{\gamma}\left(\Tr[\rho\rho_{\rm prod}]-\gamma\right)^2,\label{eq:condition-3}\\
{\rm where}\;\gamma&:=\Tr[\rho_{\rm prod}^2]-\beta^2=\Tr[\rho_{\rm prod}^2]-2^{2-m}\lambda_{\rm min}^2(\rho_{\rm prod})>0
\end{align}
Clearly, the bound given in Eq.~\ref{eq:condition-3} is larger than $\beta^2$ for the separable ball criterion, albeit the bound depends on $\rho$. Therefore, our new criterion in Eq.~\ref{eq:condition-2} implies Theorem 1. In the case that $\rho_{\rm prod}$ is the maximally mixed state $\frac{1}{d}\mathbb{I}$, then Eq.~\ref{eq:condition-2} reduces to 
\begin{align}
    \Tr[\rho^2]\leq \frac{1}{d-2^{2-m}} \Longleftrightarrow ||\rho-\rho_{\rm prod}||_F^2\leq \frac{2^{2-m}}{(d-2^{2-m})d} \label{eq:maximally-mixed}
\end{align}
which was found in the Proposition 6 of \cite{03Gurvits_Multipartite_Mixed}. In the bipartite case, the trace condition (the left inequality) in Eq.~\ref{eq:maximally-mixed} reduces to the widely-known separability condition $\Tr[\rho^2]\leq \frac{1}{d-1}$ \cite{99Braunstein_Separable_Mutipartite,02GurvitsLSBaMMBS}, while the norm condition (the right inequality) in Eq.~\ref{eq:maximally-mixed} reduces to $||\rho-\rho_{\rm prod}||_F\leq 1/\sqrt{(d-1)d}$, which indicates the largest separable ball around $\frac{1}{d}\mathbb{I}$ in the bipartite case \cite{02GurvitsLSBaMMBS}. Therefore, Eq.~\ref{eq:condition-2} generalizes the largest separable ball around the bipartite $\frac{1}{d}\mathbb{I}$ to any full-rank multipartite product state.

\section{Separable Balls around any Separable States}\label{sec:general-ball}
Based on the results in Sec.~\ref{sec:prod-ball}, we now try to characterize the existence and possible sizes of separable balls around a generic separable quantum state. For any separable state $\rho$ with ${\rm det}(\rho)=0$, that is $\rho$ is not full-rank, then $\rho\in\partial S$ according to the Corollary 4.3 of \cite{02Pittenger}. Any non-full rank separable state lies on the boundary of the set of separable states, so there exist no separable balls around these states. We next discuss the full-rank case.

Using the separable balls around the full-rank product states given in Theorem 1, we can prove the existence and characterize the size of the separable balls around a large class of full-rank separable states. We show the following result formulated in the space of Hermitian matrices.
\medskip\newline
\noindent
{\bf Corollary 2}: Suppose a separable Hermitian matrix $K$ can be decomposed into the sum of positive product matrices, that is $K=\sum_{i=1}^u K_i$ where $K_i$ are positive (semi-)definite, product matrices such that $K_i=K_{i,1}\otimes...\otimes K_{i,m}$. Suppose at least one of $\{K_i\}$ is full-rank. Then for any Hermitian matrix $M$ such that $||M-K||_F\leq 2^{1-m/2}{\rm max}_i[\lambda_{\rm min}(K_i)]$, $M$ is separable.\medskip
\newline\noindent
{\bf Proof of Corollary 2:} Let $\Delta:=M-K$, and suppose ${\rm max}_i[\lambda_{\rm min}(K_i)]$ reaches the maximum for some positive-definite product matrix $K_v$ where $1\leq v\leq u$. We then have $||\Delta||_F\leq 2^{1-m/2}\lambda_{\rm min}(K_v)$. Since at least one of $\{K_i\}$ is full-rank, then $K_v$ is full-rank and $\lambda_{\rm min}(K_v)>0$. Using Theorem 1, we know $\Delta+K_v$ is separable. Since
\begin{align}
M=K+\Delta=\sum_{i\neq v}^u K^i+(K_v+\Delta),
\end{align}
which is the sum of separable matrices, so $M$ is separable.
\medskip\newline\noindent
Considering the space of density matrices, the above corollary tells us that for any separable density matrix $\rho$ with a separable decomposition $\rho=\sum_{i=1}^u p_i\rho_i$ where $\{\rho_i\}$ are product states and at least one of $\{\rho_i\}$ is full-rank, there exists a separable ball of radius $\phi:=2^{1-m/2}{\rm max}_i[p_i\lambda_{\rm min}(\rho_i)]$ around $\rho$ in the distance induced by the Frobenius norm, and these states therefore lie in the interior of $S$. Since the separable decomposition of $\rho$ is not unique, we can then give a better lower bound $r(\rho)$ for the radius of the separable ball around $\rho$:
\begin{equation}
r(\rho)=2^{1-m/2}{\rm max}\left\{{\rm max}_i[p_i\lambda_{\rm min}(\rho_i)]:\forall\{p_i,\rho_i\}{\rm\;such\;that\;}\rho=\sum_i p_i\rho_i\right\}\label{eq:radius-general},
\end{equation}
which is the maximum value of $2^{1-m/2}{\rm max}_i[p_i\lambda_{\rm min}(\rho_i)]$ over all possible separable decompositions of $\rho$. In practice, the exact value of Eq.~\ref{eq:radius-general} is hard to determine since it is challenging to find all, if any, separable decompositions of an arbitrary separable state. 

We have therefore proved the existence and given a lower bound for the separable ball around any separable state $\rho$ that has a separable decomposition with at least one full-rank product state component. We expect most of the full-rank separable states to have such a decomposition. However, for any full-rank separable state that can only be decomposed as a convex combination of non-full rank product states, Corollary 2 no longer applies. By intuition, we expect some of these states to be on the boundary $\partial S$ while others remaining in the interior of $S$ with a separable ball around them, which reflects the complex geometric structures of $\partial S$ \cite{06Ioannou_polytrope,07Guhne_noface,15Chen_boundary}. We leave the study of these separable states that are beyond the scope of Corollary 2 to future works.

\section{Implication on Entanglement Dynamics}\label{sec:dynamics}

In the context of the dynamics of interactions, consider two finite-dimensional systems $A$ and $B$ with a total dimension $d=d_Ad_B$ and an initial state $\rho_A\otimes\rho_B$ such that ${\rm det}(\rho_A\otimes\rho_B)\neq 0$. Using perturbative methods, Ref.~\cite{22Wen} has found that the entanglement monotone negativity\footnote{Regarding the relevant properties of the entanglement monotone negativity, see \cite{02EMNegativity}.} will remain zero for a finite amount of time at the onset of the interaction, regardless of the interaction Hamiltonians between the two systems. This is because the eigenvalues of the partial transpose of the density matrix all start out positive (as ${\rm det}(\rho_A\otimes\rho_B)\neq 0$), and it will, therefore, take a finite amount of time for at least one eigenvalue to become negative and thus contribute to negativity. When the total dimension of the bipartite systems $d\leq 6$, the vanishing negativity does imply separability according to the PPT criterion \cite{96PeresSeparability}, which implies $A$ and $B$ will remain unentangled for a finite amount of time as discussed by \cite{22Wen}. However, vanishing negativity is itself not a sufficient criterion for separability, i.e., for a general initial state $\rho_A\otimes\rho_B$ with an arbitrary finite dimension, we can not conclude $A$ and $B$ will remain unentangled using the framework in \cite{22Wen}.

Equipped with the bipartite version of Theorem~1 in this work, we can now strengthen the result of vanishing negativity in \cite{22Wen} and directly conclude that the entanglement between $A$ and $B$ can only kick in after a finite amount of time for any full-rank initial state $\rho_A\otimes\rho_B$. As the time evolution of the state is continuous, it will take a finite amount of time for $\rho_{AB}(t)$ to evolve beyond the boundary of the separable ball around the positive-definite $\rho_A\otimes\rho_B$ with a minimum radius of $\lambda_{\rm min}(\rho_A\otimes\rho_B)$. Therefore, no interaction Hamiltonians can immediately entangle $A$ and $B$ at the onset. The two systems are guaranteed to remain unentangled for some time after two systems start to interact. We can further generalize the above argument to the multipartite case: for any quantum systems with an initial state $\rho_{\rm prod}=\rho_1\ot...\ot\rho_m$ such that ${\rm det}(\rho_{\rm prod})\neq 0$, these $m$ subsystems will remain unentangled for a finite amount of time regardless of the Hamiltonian, due to the existence of a separable ball around the initial state. 

Using the radius of the separable ball $\lambda_{\rm min}(\rho_A\otimes\rho_B)$, we can in fact perturbatively estimate the lower bound for the amount of time it takes to entangle $A$ and $B$. For a bipartite initial state $\rho_0=\rho_A\otimes\rho_B$ (${\rm det}(\rho_A\otimes\rho_B)\neq 0$) and a time-independent interaction Hamiltonian $H_{\rm int}=\sum_p A_p\otimes B_p$ acting on on $A$ and $B$, the time evolution of the density matrix is given as  
\begin{align}
\rho(t)=\rho_A\otimes\rho_B+it[\rho_A\otimes\rho_B,H_{\rm int}]+O(t^2),
\end{align}
upto the first order. Ignoring the second-order, we have then 
\begin{align}
||\rho(t)-\rho_A\otimes\rho_B||_F\approx t\left\lvert\left\lvert[\rho_A\otimes\rho_B,H_{\rm int}]\right\rvert\right\rvert_{F}
\end{align}
Let the amount of time it takes for $H_{\rm int}$ to entangle $A$ and $B$ from the start of interaction be $t_{\rm E}$. For $\rho(t)$ to become entangled, we need the state to evolve beyond the separable ball around $\rho_A\otimes\rho_B$ such that $||\rho(t)-\rho_A\otimes\rho_B||_F>\lambda_{\rm min}(\rho_A\otimes\rho_B)$, then within the first-order perturbation theory, $t_{\rm E}>\lambda_{\rm min}(\rho_A\otimes\rho_B)/\left\lvert\left\lvert[\rho_A\otimes\rho_B,H_{\rm int}]\right\rvert\right\rvert_{F}$, which is a lower estimate for the amount of time it takes for $\rho(t)$ to become entangled under $H_{\rm int}$. Note that for a positive semi-definite, diagonal matrix $D$ and an arbitrary matrix $U$, we have
\begin{align}
    ||[D,U]||_F^2&=||DU-UD||_F^2=\sum_{ij}|(D_{ii}U_{ij}-U_{ij}D_{jj})|^2\nonumber\\
    &=\sum_{ij}|(D_{ii}-D_{jj})|^2|U_{ij}|^2\leq ({\rm max}(D_{ii})-{\rm min}(D_{ii}))^2 ||U||_F^2
\end{align}
Therefore, working in the basis of $\rho_0=\rho_A\otimes\rho_B$ such that the initial density matrix is diagonal, we have
\begin{align}
t_{\rm E}&>\frac{\lambda_{\rm min}(\rho_0)}{\left\lvert\left\lvert[\rho_0,H_{\rm int}]\right\rvert\right\rvert_{F}}\geq \frac{\lambda_{\rm min}(\rho_0)}{\lambda_{\rm max}(\rho_0)-\lambda_{\rm min}(\rho_0)}\frac{1}{\left\lvert\left\lvert H_{\rm int}\right\rvert\right\rvert_{F}}\label{eq:estimate-time}\\
\Longleftrightarrow t_{\rm E} \left\lvert\left\lvert H_{\rm int}\right\rvert\right\rvert_{F}&>\frac{\lambda_{\rm min}(\rho_0)}{\lambda_{\rm max}(\rho_0)-\lambda_{\rm min}(\rho_0)}\label{eq:uncertainty-time}.
\end{align}
When $\lambda_{\rm min}(\rho_0)=\lambda_{\rm max}(\rho_0)$, that is $\rho_0=\frac{1}{d}\mathbb{I}$ is maximally mixed, the bound becomes infinity, suggesting that $A$ and $B$ will never become entangled, which is indeed the case since $\frac{1}{d}\mathbb{I}$ is invariant under unitary evolution. When $\lambda_{\rm min}(\rho_0)=0$, that is $\rho$ is not full-rank, the bound in Eq.~\ref{eq:estimate-time} reduces to 0, which is consistent with the fact that the state can be immediately entangled by $H_{\rm int}$. Interestingly, Eq.~\ref{eq:uncertainty-time} takes a form similar to the time-energy uncertainty principle. In general, Eq.~\ref{eq:estimate-time} gives an approximate lower bound for the amount of time it takes to entangle bipartite product states under any interaction Hamiltonian, which can be of interest to quantum experiments aiming to create entanglement or protect separability. 

\section{Conclusions and Outlook}\label{sec:conclusion}
In this work, we extend the results of \cite{02GurvitsLSBaMMBS,03Gurvits_Multipartite_Mixed,20Hanson_Markovian} by proving the existence of a separable ball around any full-rank multipartite product state $\rho_{\rm prod}=\rho_1\otimes...\otimes\rho_m$ in Sec.~\ref{sec:prod-ball}. We further establish a lower bound for the radius of the ball in terms of distance induced by the Frobenius norm, given by $\beta=2^{1-m/2}\prod_{i=1}^m\lambda_{\rm min}(\rho_i)$. In addition, we also give the bound in terms of distance induced by other Schatten $p-$norms and quantum relative entropy. In the bipartite case, we show that the radius $\lambda_{\rm min}(\rho_1)\lambda_{\rm min}(\rho_2)$ is the largest in the space of Hermitian matrices. While in the space of density matrices this radius is generally not the maximum radius, it is close to the maximum radius for the high dimensional bipartite systems. In further work, it should be possible to determine the maximum radii for the separable balls found in this work. This task is likely to require the further use of properties unique to density matrices, such as the unit trace and positivity.

Applying a scaling relation to Theorem~1, we obtain a new sufficient criterion for multipartite separability in Sec.~\ref{sec:criterion}: if $\rho$ satisfies $\Tr[\rho\rho_{\rm prod}]^2/\Tr[\rho^2]\geq \Tr[\rho_{\rm prod}^2]-\beta^2$ where $\beta=2^{1-m/2}\lambda_{\rm min}(\rho_{\rm prod})$ for any product state $\rho_{\rm prod}$, then $\rho$ is separable. This criterion is easy to compute, and it reduces to the largest separable ball condition around the bipartite maximally mixed state if we set $\rho_{\rm prod}=\frac{1}{d}\mathbb{I}$. To fully exploit this criterion for showing the separability of an arbitrary state $\rho$, we need to be able to find an $m$-partite product state $\rho_{\rm prod}$ that is suitably close to $\rho$ in the Frobenius metric or some other metric, which requires further theoretical and numerical study. For example, for a given $\rho$, an intuitive place to start is to consider the tensor product of the reduced density matrices, that is $\Tr_{1}[\rho]\otimes...\otimes\Tr_{m}[\rho]$. Alternatively, one can try to find the closest $m$-partite product states in a given metric. For the bipartite version of the problem of finding the closest product state in the Frobenius norm, see the appendix of \cite{02Lockhart_Product_State} for a numerical algorithm. 

Using the separable balls around the full-rank product states, we show the existence of a separable ball around any separable state $\rho$ that has a separable decomposition with at least one full-rank product state component, that is $\rho=\sum_{i=1}^u p_i\rho_i$ where $\{\rho_i\}$ are product states and at least one of $\{\rho_i\}$ is full-rank. We give an explicit lower bound for the radius of the separable ball around $\rho$ as $2^{1-m/2}{\rm max}_i[p_i\lambda_{\rm min}(\rho_i)]$. 
This result significantly expands the past literature work on separable balls around the maximally mixed state to the existence and possible sizes of the separable ball around a generic separable state.

Our finding of a separable ball around any fully mixed multipartite product state sheds new light on entanglement dynamics in interactions, which can be valuable for applications in quantum technologies. In particular, our results imply that any multipartite quantum system beginning with a fully mixed product state will remain unentangled for a finite amount of time, regardless of the Hamiltonian of the system. In the bipartite case, we also estimate the amount of time it takes for any interaction Hamiltonian to entangle two particles initially at a product state, which is given in Eq.~\ref{eq:estimate-time}. 

Finally, we note that our work may be useful for studies of the relationship between $(n+1)$-partite entanglement and $n$-partite entanglement, as any mixed $n$-partite state can be purified by appending a sufficiently large ancilla system. For instance, our work could inform studies of tri-partite entanglement \cite{19CunhaTripartiteEntanglement}.

\section*{References}
\bibliographystyle{iopart-num}  
\bibliography{refs}

\end{document}